\documentclass[aps,prl,preprint,superscriptaddress,floatfix]{revtex4-1}
\usepackage{amsmath}
\usepackage{amsfonts} 
\usepackage{graphicx}
\usepackage{verbatim}
\usepackage{natbib}
\usepackage{threeparttable}
\usepackage{xcolor}
                                           
\setcitestyle{super}

\begin{document}

\begin{center}

\textbf{\large Ultrastrong Light-Matter Coupling in Materials}\\[0.5em]

Niclas S. Mueller,$^{1,2}$ Eduardo B. Barros,$^2$ and Stephanie Reich$^{1,*}$\\[0.5em]

\textit{$^1$ Department of Physics, Freie Universit\"at Berlin,\\ Arnimallee 14, D-14195 Berlin, Germany\\}

\textit{$^2$ Department of Physical Chemistry, Fritz Haber Institute of the Max Planck Society, D-14195 Berlin, Germany\\}

\textit{$^3$ Department of Physics, Universidade Federal do Cear\'a, Fortaleza,\\ Cear\'a, 60455-760 Brazil\\}

\end{center}
\vspace{1cm}

Ultrastrong light-matter coupling has traditionally been studied in optical cavities, where it occurs when the light-matter coupling strength reaches a significant fraction of the transition frequency. This regime fundamentally alters the ground and excited states of the particle-cavity system, unlocking new ways to control its physics and chemistry. However, achieving ultrastrong coupling in engineered cavities remains a major challenge.  Here, we show that ultra- and deep-strong coupling naturally occur in bulk materials without the need for external cavities.  By analyzing experimental data from over 70 materials, we demonstrate that phonon-, exciton-, and plasmon-polaritons in many solids exhibit ultrastrong coupling, systematically surpassing the coupling strengths achieved in cavity-based systems. To explain this phenomenon, we introduce a dipole lattice model based on a generalized Hopfield Hamiltonian, which unifies photon-matter, matter-matter, and photon-photon interactions. The complete overlap between the photonic and collective dipole modes in the lattice enables ultrastrong coupling, leading to excited-state mixing, radiative decay suppression, and potential phase transitions into collective ground states. Applying our model to real materials, we show that it reproduces light-matter coupling across broad material classes and may underlie structural phase transitions that give rise to emergent phenomena such as ferroelectricity, insulator-to-metal transitions, and exciton condensation. Recognizing ultrastrong coupling as an intrinsic property of solids reshapes our understanding of light-matter interactions and opens new avenues for exploring quantum materials and exotic phases of matter.

\newpage
\section{Introduction}

Cavity quantum electrodynamics (QED) has long explored the idea of light-matter (LM) coupling shaping particle states.\cite{Kaluzny1983,Rempe1993,Reithmaier2004} A particle in a photonic cavity may be understood as an artificial molecule with hybridized cavity-particle excitation: LM coupling splits the initially degenerate cavity photon $\omega_\mathrm{pt}$ and particle excitation $\omega_0$ into two polaritons that are separated by twice their coupling strength $g$. This splitting becomes observable experimentally in the strong coupling regime where $g$ exceeds the system losses $\gamma$.\cite{Rempe1993,Hood1998,Reithmaier2004}
Strong coupling changes the nature and properties of the excited states, but the ground state of the cavity-particle system remains unaffected. With further increasing LM coupling, however, the ground state becomes populated with virtual photons and particle excitations 
gaining energy in the ultrastrong (USC, $g\gtrsim 0.1\omega_0$) and deep strong (DSC, $g>\omega_0$) coupling regimes.\cite{Ciuti2005,Mueller2020,Kockum2019,FornDiaz2019} Fascinating physics has been predicted for USC cavity molecules like squeezed photons, deterministic non-linear optics with single photons, electroluminescence from the ground state, and phase transitions into novel states.\cite{Kockum2019,FornDiaz2019} These systems have been explored for cavity control of chemical reactions, ferroelectricity, and superconductivity, where USC  shifts the system eigenstates.\cite{Vidal2021, Schlawin2022, Martinez2018, Ruggenthaler2023, Ashida2020, Latini2021, Curtis2023}  

Concepts of ultrastrong LM coupling have rarely been applied to solids, although certain three-dimensional (3D) materials are striking examples of ultrastrong and deep strong LM coupling without needing external cavities. In barium titanate the LM interaction strength $g$ of the phonon-polaritons exceeds the transverse phonon frequency $\omega_\mathrm T$ by a factor of two or more.\cite{Luspin1980} Another example are nanoparticle supercrystals that were recently shown to have a coupling strength of several eV.\cite{Mueller2020} Many materials have a polaritonic band gap or fully-reflective Reststrahlenband in their optical spectra -- a fingerprint of the USC regime -- giving rise to surface polaritons and volume-confined hyperbolic polaritons with sub-wavelength confinement, nanoscale waveguiding, and hyperlensing.\cite{Galiffi2024, Wang2024, Kowalski2025} Despite clear evidence of USC in solids and the use of  polaritons in interpreting their optical spectra,\cite{Hopfield1958,YuCardona,ClaudioAndreani1995} LM coupling has been neglected in models of crystal ground states. This raises questions about the pervasiveness of USC and DSC in 3D materials and the extent to which LM coupling influences their physics and chemistry.\cite{Canales2021,Kockum2019,FornDiaz2019} 

Here we highlight the prevalence of USC and DSC in bulk materials and how LM coupling shapes their ground and excited states. Analyzing a large set of available experimental data, we show that phonons, excitons, and plasmons in many materials fall into the USC and DSC regimes, exceeding the LM coupling of the same material in photonic and plasmonic cavities. To understand USC in solids, we set up a lattice model of transition dipoles that is described by a generalized Hopfield Hamiltonian including dipole-dipole interactions. 
Its ground- and excited states show how  photon-matter, matter-matter, and photon-photon interaction govern the lattice response. In selected material classes, we discuss how LM coupling deeply affects their optical, vibrational, and mechanical properties. Realizing the fundamental role of LM coupling for solids will help to tailor them, e.g., for polaritonic and ferroelectric devices, preparing nonclassical states of light, and guiding the search for materials with exotic ground states. 

\section{Results and Discussion}

\subsection{Evidence for USC in Bulk Materials}

\begin{figure*}[h!]
    \includegraphics[width=16.4cm]{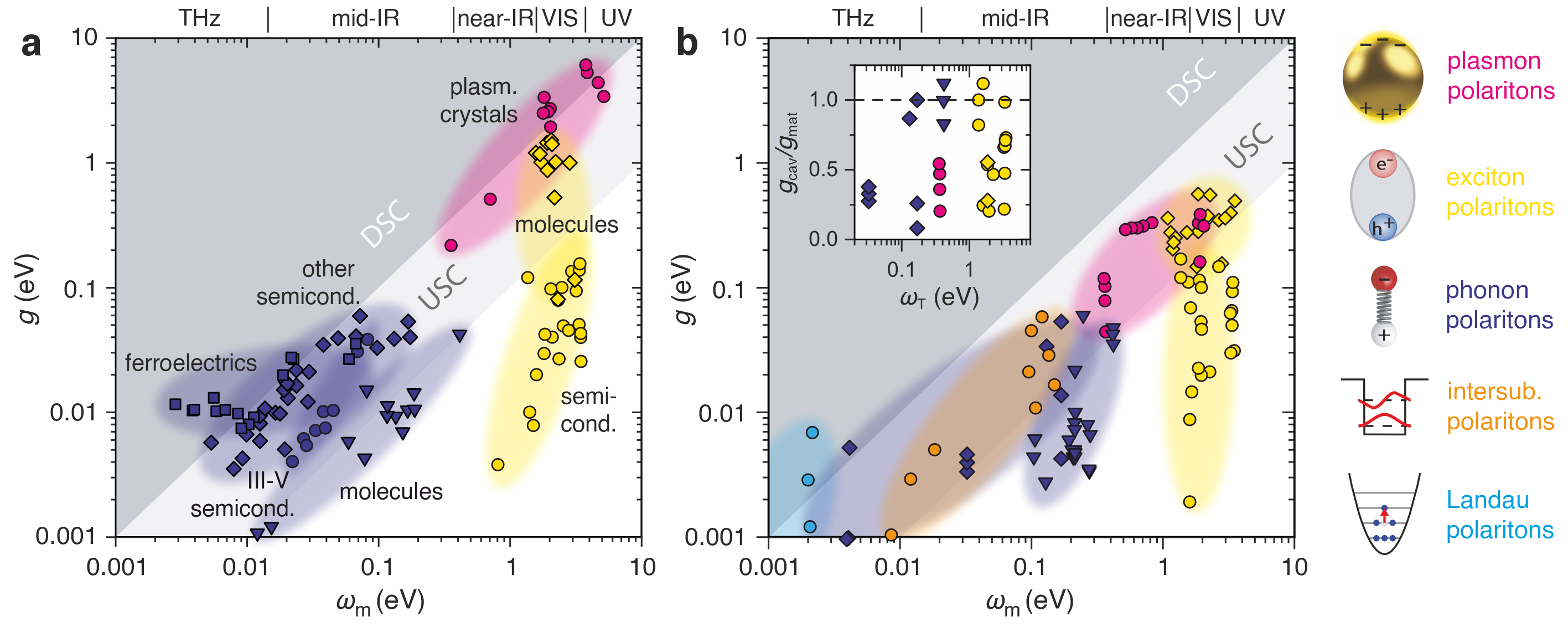}
    \caption{USC and DSC in bulk and cavity polaritons. (a) Coupling constant  $g$ versus frequency $\omega_\mathrm{m}$ for (a) bulk polaritons and (b) materials in cavities. Blue symbols are for phonon-, yellow for exciton-, and pink for plasmon-polaritons. Intersubband (orange) and Landau polaritons (turquoise) in (b) are inherently cavity excitations without bulk counterparts. The inset in (b) shows that light-matter coupling in cavities $g_\mathrm{cav}$ remains below the bulk limit $g_\mathrm{mat}$ for phonons (blue), excitons (yellow), and plasmons (pink). The Supplementary Tables~S.3-17 contain the data shown in (a,b). The light (dark) shaded areas represent the USC (DSC) regime.}
    \label{fig:mat}
\end{figure*}

 Ultrastrong LM coupling occurs when the coupling strength becomes comparable to the frequency of an excitation, regardless of losses, see shaded areas in Fig.~\ref{fig:mat}. Achieving this in photonic cavities is challenging, where it typically requires many particles interacting with the photonic mode.\cite{Kockum2019,FornDiaz2019} Here we show that USC is common in bulk polaritons, where most materials exhibit at least one type of excitation -- phonon, exciton, or plasmon -- within the USC or DSC regime. To explore this, we re-analyzed the optical spectra of over 70 materials extracting their excitation frequency $\omega_m$ and LM coupling strength $g$, see Fig.~\ref{fig:mat}a, Supplementary Tables~S3-9, and Methods. We found that $g$ increases linearly with $\omega_m$ over four orders of magnitude from a few meV for low-frequency phonons (purple) to several eV for plasmon-polaritons (pink).  Most of the analyzed excitations fall within the USC regime $g \gtrsim 0.1\omega_\mathrm{m}$,  light shaded area in Fig.~\ref{fig:mat}a. This includes infrared-active phonons in partially ionic semiconductors and organic crystals (see labels) as well as excitons (yellow) in organic crystals and two-dimensional materials. Deep strong coupling ($g>\omega_\mathrm{m}$, dark shaded areas) is observed for plasmon-polaritons in supercrystals and  phonon-polaritons in ferroelectrics. Interestingly, traditional semiconductor excitons (yellow circles) are a notable exception to USC, with $g\sim 10^{-2}\omega_\mathrm{m}$. This is particularly striking given that III-V semiconductors, such as gallium arsenide have been prime materials for  strong LM coupling due to their superior processability, high crystal quality, and low losses. Figure~\ref{fig:mat} may guide the selection of the most suitable materials to study USC and DSC phenomena.

Light-matter coupling of bulk polaritons is systematically larger than for the same material confined within photonic and plasmonic cavities, Fig.~\ref{fig:mat}b, although cavities are commonly perceived as key tools to increase LM  interaction. Figure \ref{fig:mat}b summarises data from $>$80 cavity-based setups and demonstrates that their  $g$ are consistently lower than those observed in bulk material classes in Fig.\ \ref{fig:mat}a.  This limit has been experimentally confirmed for phonon-, exciton-, and plasmon polaritons in planar microcavities,
inset in Fig.~\ref{fig:mat}b,\cite{Barraburillo2021,Dirnberger2023,Chang2025} but it also applies to plasmonic cavities, metamaterials, and  other photonic devices, see Supplementary Table S17. While the experimental trends clearly establish the prevalence of USC in bulk materials, we need a quantitative framework to explain why materials naturally enter the USC and DSC regimes. For this, we develop a microscopic lattice model of solid-state material excitations.

\subsection{The Dipole Lattice Model: A Unified Theory for Bulk Polaritons }

\begin{figure*}
    \includegraphics[width=14cm]{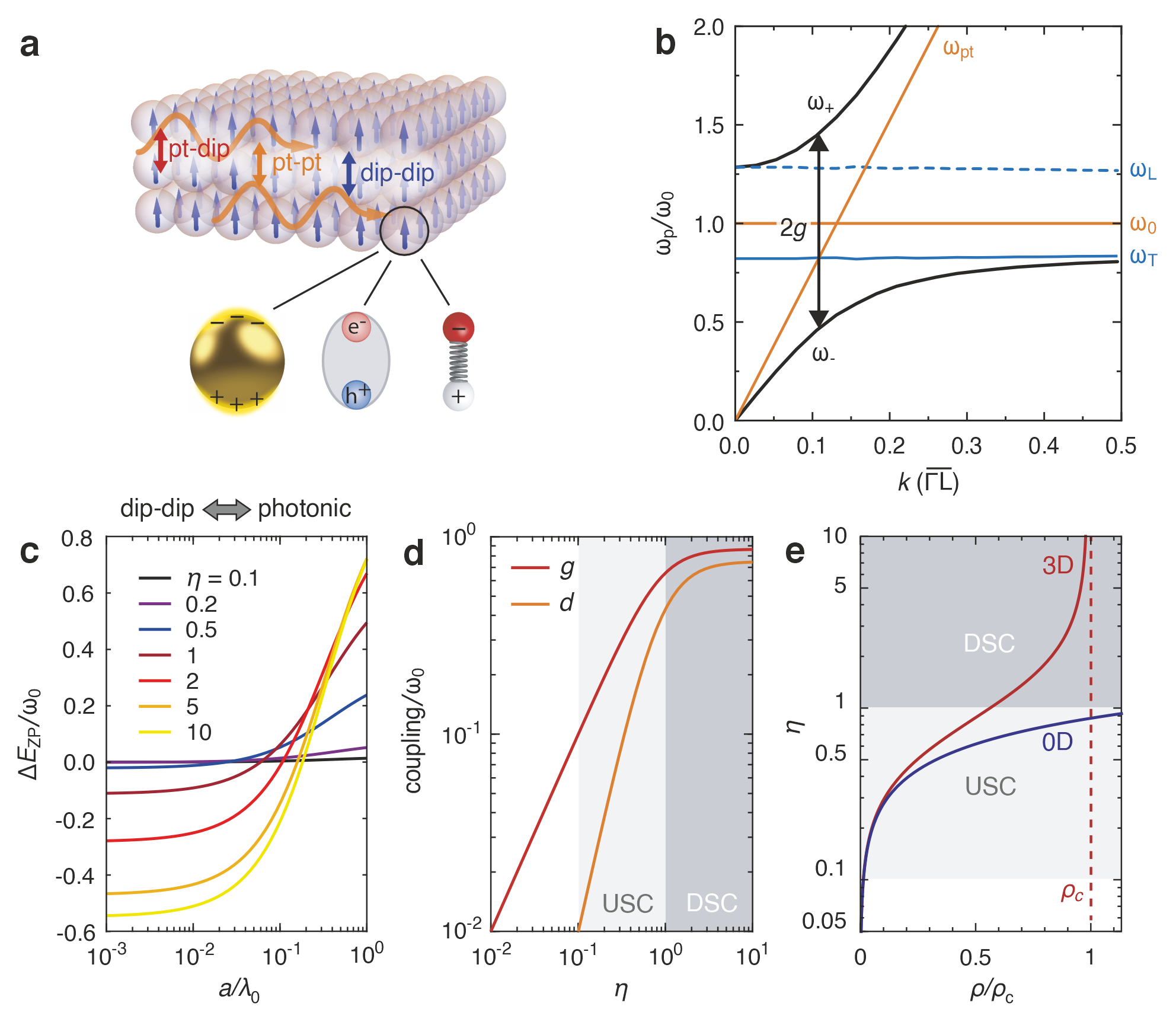}
     \caption{Dipole lattice model and ultrastrong LM coupling. (a) Lattice of transition dipoles (blue arrows, spheres show $V_\mathrm{dip}$) representing plasmons, excitons, and vibrations (lower row). The double arrows indicate photon-dipole (pt-dip), photon-photon (pt-pt), and dipole-dipole (dip-dip) coupling. (b) Lower $\omega_-$ and upper $\omega_+$ polariton (black lines) of an fcc dipole lattice. Orange: photon $\omega_\mathrm{pt}$ and bare dipole $\omega_0$ dispersion, blue line: collective transverse $\omega_\mathrm{T}$ (full) and longitudinal mode $\omega_{\mathrm{L}}$ (dashed), ($f/V_\mathrm{dip} = 2\sqrt{2}\,\mathrm{nm}^{-3}$, 
 $\hbar \omega_0 = 2$\,eV, $\epsilon_\infty = 1$, and $m = m_e$). (c) $\Delta E_\mathrm{ZP}$ as a function of the ratio between the lattice constant $a$ and the vacuum wavelength of the dipole $\lambda_0 = 2\pi c/\omega_0$ for various $\eta$. 
 (d) Photon-dipole $g$ and dipole-dipole coupling strength $d$ as a function of $\eta$. 
 (e) $\eta$ as a function of density $\rho = f/V_\mathrm{dip}$ for a 3D fcc lattice  (red line) and for a single dipole in a cavity (blue). }
    \label{fig:Fig3}
\end{figure*}

To understand the widespread occurrence of USC in bulk materials and its implications for both ground and excited states, we developed a microscopic model of material excitations and their interaction with photons. The model describes bulk polaritons within the framework of cavity QED, treating them as eigenstates of a lattice of transition dipoles, Fig.~\ref{fig:Fig3}a. This general approach provides a unified description of optically active excitations in many material platforms. Conceptually, the model can be viewed as a generalized Hopfield Hamiltonian, incorporating dipole-dipole interactions between localized excitations alongside photon-dipole and photon-photon interactions, arrows in Fig.~\ref{fig:Fig3}a.\cite{Kockum2019,FornDiaz2019,Hopfield1958,Lamberto2024,muniain2024} 
Essentially, it condenses the electronic and ionic response of a material into its optically active excitations, represented as a lattice of locally induced transition dipoles, Fig.~\ref{fig:Fig3}a, with frequency $\omega_0$, oscillator strength $f$, mass $m$, and volume $V_\mathrm{dip}$.\cite{Lamowski2018,Barros2021} 

The transition dipoles in our model may represent excitations of structural building blocks, as is the case for nanoparticle supercrystals and organic crystals. However, in traditional solids such as metals, covalent compounds, and ionic materials, where the system cannot be decomposed into distinct monomeric units, the definitions of $\omega_0$ and $V_\mathrm{dip}$ become more abstract and do not necessarily correspond to real transitions.
For example, for a Wannier exciton of a semiconductor, $\omega_0$ represents the bound state formed by the direct Coulomb interaction between an electron and a hole and $V_\mathrm{dip}$ is determined by the Bohr radius of this exciton. In the case of  phonons, $V_\mathrm{dip}$ corresponds to the volume occupied by vibrating atoms within a formula unit and $\omega_0$ represents their vibrational frequency induced by covalent and ionic bonding in the absence of dipole coupling between units. Although $\omega_0$ is technically momentum-dependent in 3D materials due to quasiparticle dispersion, we omit this dependence for simplicity. Importantly, the model is not limited to single bosonic excitations but can be extended to higher-order multipoles, fermionic excitations, and multiple excitation frequencies.\cite{DavydovBook, Barros2021}

Each transition dipole in Fig.~\ref{fig:Fig3}a interacts with the other dipoles via Coulomb coupling (dip-dip) and couples to propagating photons (pt-dip) inside the crystal. The resulting generalized Hopfield Hamiltonian intricately intertwines dipole-dipole, photon-dipole, and photon-photon interactions, \cite{Rempe1993,Kockum2019,FornDiaz2019,Todorov2012,Lamberto2024,muniain2024} see Fig.~\ref{fig:Fig3}a and Methods for further discussion,
\begin{equation}
\label{eq:LMHam}
\begin{split}
\mathcal{H} = &\  \hbar \omega_0 b_\mathbf{k}^\dagger  b_\mathbf{k} + \hbar \omega_{\mathrm{pt},\mathbf{k}} a_\mathbf{k}^\dagger  a_\mathbf{k} 
\\ & + \hbar  d s_\mathbf{k}(b_\mathbf{k}^\dagger + b_\mathbf{-k})(b_{-\mathbf{k}}^\dagger + b_\mathbf{k})
\\ & + i \hbar g \xi_\mathbf{k} (a_\mathbf{k}^\dagger + a_{-\mathbf{k}})(b_{-\mathbf{k}}^\dagger - b_\mathbf{k})
\\ & + \hbar \frac{g^2}{\omega_0} \xi_\mathbf{k}^2 (a_\mathbf{k}^\dagger + a_{-\mathbf{k}})(a_{-\mathbf{k}}^\dagger + a_\mathbf{k}).
\end{split}
\end{equation}
The creation and annihilation operators act on the dipole ($b^\dagger$, $b$) and photon ($a^\dagger$, $a$) modes; $\mathbf{k}$ is the wavevector. The first two terms in Eq.~\eqref{eq:LMHam} describe the uncoupled dipoles $\omega_0$ and photons $\omega_{\mathrm{pt},\mathbf{k}}$, orange lines in Fig.~\ref{fig:Fig3}b. The third term encodes dipole-dipole interaction that couples the localized dipoles into collective transverse $\omega_\mathrm{T}$ and longitudinal $\omega_\mathrm{L}$ states controlled by the coupling constant $d$ and the lattice sum $s_\mathbf{k}$, full and  dashed blue lines in Fig.~\ref{fig:Fig3}b. The fourth term represents LM interaction with propagating photons inside the crystal; it generates polaritonic states depending on   $g$ and the detuning between the photon and the bare dipole frequency $\xi_\mathbf{k}=\sqrt{\omega_0/\omega_{\mathrm{pt},\mathbf{k}}}$. The final $A^2$ term accounts for photon-photon interaction, which is crucial in the USC regime.\cite{Hopfield1958}

The polariton dispersion and the ground state of the dipole lattice are determined by LM coupling. The collective excitations of the system are the lower $\omega_{\mathbf{k}, -}$ and upper polariton $\omega_{ \mathbf{k},+}$, Fig.~\ref{fig:Fig3}b black lines, with a minimum separation given by the Rabi splitting $\Omega_R=2g$; $g$ corresponds to the vacuum Rabi frequency in cavity QED. The dispersion  resembles the polaritonic response of a single dipole $\omega_0$, but the collective frequency $\omega_\mathrm T\approx \sqrt{\omega_0^2-4d\omega_0/3}$ is downshifted compared to $\omega_0$ by the dipole-dipole coupling.\cite{DavydovBook,Lamowski2018,Barros2021} Consequently, the reduced coupling strength is given by $\eta=g/\omega_\mathrm T$.

USC and DSC shift the ground state energy of the dipole lattice. The LM contribution to the zero point energy is given by the energy difference between coupled and uncoupled excitations,\cite{Ciuti2005,Quattropani1986,Kockum2019,FornDiaz2019} see Methods,
\begin{equation}
\label{eq:BGE0} 
\begin{split}
\Delta E_\mathrm{ZP}= \frac{1}{2N_k}\sum_{\mathbf k} 
\Big[
\sum_\sigma \left( \hbar \omega_{\mathbf{k}, \sigma, -} +  \hbar \omega_{\mathbf{k}, \sigma,+}\right) + \hbar \omega_{\mathrm{L},\mathbf{k}} -  3 \hbar \omega_0
- 2 \hbar \omega_{\mathrm{pt},\mathbf k}
\Big]
\end{split}
\end{equation}
summed over all polarizations $\sigma$ and averaged over $N_k$ wavevectors $\mathbf{k}$ in the first Brillouin zone. The shift becomes significant in the USC regime for $\eta\gtrsim 0.5$, with both its magnitude and sign depending on  $\eta$ and the translational periodicity of the dipole lattice, Fig.\,\ref{fig:Fig3}c. When the unit cell is much smaller than the vacuum wavelength of the bare transition $\lambda_0=2\pi c/\omega_0$, the ground state energy decreases with increasing $\eta$ due to dipole-dipole coupling, Fig.~\ref{fig:Fig3}c. Conversely, when the lattice constant $a$ is comparable to or larger than $\lambda_0$ -- as seen in metamaterials and artificial supercrystals -- the total energy increases and the lattice is dominated by photonic contributions to its ground state, Fig.~\ref{fig:Fig3}c. 

The generalized Hopfield Hamiltonian provides an analytic expression for the LM coupling strength of the dipole lattice, see Methods,
\begin{equation}
    \label{eq:g3D}
    g = \sqrt{\frac{e^2 f}{4\epsilon_0 \epsilon_\infty m V_\mathrm{dip}}},
\end{equation}
where we introduced the high-frequency dielectric constant $\epsilon_\infty$ to account for other excitations in addition to $\omega_0$, see Methods. This expression closely resembles the formula for LM coupling of a single dipole in a cavity, with the key distinction that $V_\mathrm{dip}$ replaced the photonic mode volume $V_\mathrm{cav}$.\cite{Kockum2019,FornDiaz2019,Canales2021} 
Since material excitations in a 3D lattice occupy a much smaller volume than a confined photon mode, LM coupling in a dipole lattice can greatly exceed that of a single dipole in a cavity. For example, in a system of oscillating electrons ($f=1, m=m_e$) using a characteristic volume $V_\mathrm{dip}\approx 10^{-2}$\,nm$^3$ for an atom in a crystal lattice, we find an an upper bound  $g \approx 3\,$eV [this and all following estimates assume a face-centered cubic (fcc) lattice, see Methods for further details]. This is orders of magnitude larger than LM  coupling for the same dipole in photonic ($g_\mathrm{cav}\sim 0.1\,$meV) or plasmonic ($\sim10\,$meV) cavities.
In cavity-based systems, USC has been achieved experimentally by incorporating a large number of dipoles into the cavity.\cite{Kockum2019,FornDiaz2019} Only when the material completely fills $V_\mathrm{cav}$, $g_\mathrm{cav}$ reaches the limit $g$ of the 3D polariton. The generalized Hopfield model thus explains why LM coupling for bulk materials, Fig.~\ref{fig:mat}a, is inherently stronger than in the same material confined in cavities, Fig.~\ref{fig:mat}b. 

The LM (photon-dipole) coupling in the dipole lattice is intrinsically linked to its dipole-dipole interaction since the latter arises from the exchange of virtual photons. We find $d=g^2/\omega_0$, see Methods, which can also be expressed as 
\begin{equation}
\frac{d}{g}=\frac{\eta}{\sqrt{1+4\eta^2/3}}.  \label{eq:dg}  
\end{equation} 
$d$ also appears as the pre-factor of the $A^2$ term that describes the photon-photon interaction in the material. For small reduced coupling, where $d/g \approx \eta \ll1$,  dipole-dipole interaction is negligible, and LM coupling acts as a perturbation -- consistent with the common approximation for weak and strong coupling, Fig.~\ref{fig:Fig3}d. However, in the USC regime, the energy scales in Eq.~\eqref{eq:LMHam} -- the bare frequency $\omega_0$, the dipole-dipole $d$ and the photon-dipole $g$ interaction -- become comparable, Fig.\,\ref{fig:Fig3}d.
As a result, they collectively determine the eigenstates of the dipole lattice and none can be ignored or treated perturbatively.  Equation~\eqref{eq:dg} also implies that LM coupling is directly related to the energy splitting of the transverse and longitudinal collective modes $g=\sqrt{\omega_\mathrm L^2-\omega_\mathrm T^2}/2$.\cite{Barraburillo2021}

The reduced coupling strength of the dipole lattice is inherently unbound, Fig.~\ref{fig:Fig3}e, which is an important difference to a single or a few particles in a cavity. An increase in the lattice dipole density $\rho=f/V_\mathrm{dip}$ increases $g$ and $d$, which  simultaneously lowers $\omega_\mathrm T$. $\eta$ increases well into the DSC regime and  diverges at a critical dipole density $\rho_c = 3\epsilon_0\epsilon_\infty m \omega_0^2/e^2$, Fig.~\ref{fig:Fig3}e, when $\omega_\mathrm T$ approaches zero. A further increase in dipole density leads to an imaginary frequency of the transverse collective mode. This destabilizes the dipole lattice leading to a transition into a novel ground state. The behavior is reminiscent of the superradiant phase transition discussed in cavity QED, which is  prevented by the stabilizing $A^2$ term for single or non-interacting  dipoles in a cavity.\cite{Kockum2019,Pilar2020,Lamberto2024,Mueller2020} The dipole lattice, in contrast, can undergo this phase transition due to the Coulomb interaction between its transition dipoles.

\subsection{Validation of the Dipole Lattice Model for Materials}

 Before exploring the predictions of the dipole lattice model for real materials, we  demonstrate that it quantitatively reproduces the coupling strength of bulk polaritons. Table~\ref{tab:verify} shows that the predicted LM coupling strengths $g$ are in close agreement with experimental values across diverse material classes, supporting the validity of the dipole lattice model.

Organic crystals serve as an ideal test case, as the transitions of their molecular monomers directly correspond to the dipoles in the generalized Hopfield Hamiltonian. Table~\ref{tab:verify} highlights the excellent agreement between the measured and predicted $g$ for  excitons in a squarylium crystal and infrared-active vibrations in a dinitrobenzene crystal,\cite{TristaniKendra1984,Wojcik2005} see Methods and Supplementary Information. Similarly, the model predicts DSC in gold nanoparticle supercrystals, which exhibit the highest absolute Rabi frequencies among 3D materials studied so far.\cite{Mueller2020} In contrast, traditional semiconductors such as gallium arsenide and cadmium selenide experience strong dielectric screening, leading to large dipole volumes, low dipole densities, and weak LM coupling for Wannier excitons. This explains why excitons in these materials fall below the general USC trend observed in  Fig.~\ref{fig:mat}a.\cite{Hooft1987,ClaudioAndreani1995} 
Infrared-active phonons provide another class of optical excitations. In purely ionic crystals like sodium chloride, the sodium and chlorine ions carry a charge of  $|e|$, corresponding to an oscillator strength $f=1$ and $g=110\,$cm$^{-1}$, in excellent agreement with experimental values.\cite{Raunio1969} Higher oxidation states, such as in magnesium oxide ($f=2$) further enhance $g$. While the lattice dipole model describes purely ionic solids well, it underestimates $g$ in materials with mixed covalent-ionic bonding. In such cases, exciting an infrared-active vibration leads to  a redistribution of charge, causing the effective oscillator strength to exceed the static charge transfer by up to an order of magnitude.\cite{Zamponi2012} These effects are beyond the scope of our semi-empirical model, which assumes fixed charge distribution upon excitation. Capturing this fully requires a comprehensive material-light Hamiltonian that explicitly incorporates electrons, nuclei, and photons as well as their interactions. 

\begin{table*}
\caption{Predicted and measured $g$ for excitations in solids, see Methods and Suppl. Tab.\,S2.}
    \begin{tabular}{llcccc}\hline
    material &excitation& $\omega_\mathrm{T}$    &  \multicolumn{2}{c}{$g$}&  $\eta$\\
    &&&theory&exp&\\\hline
        Au supercrystals&plasmon&1.8\,eV   &  3.40\,eV   & 3.35\,eV  &  1.86\\
    squarylium &exciton&   1.59\,eV   &  1.01\,eV &  1.20\,eV &  0.75\\
    NaCl &vibration&  163\,cm$^{-1}$ &       106\,cm$^{-1}$ &    103\,cm$^{-1}$ & 0.63\\
    MgO &  vibration& 408\,cm$^{-1}$ &     212\,cm$^{-1}$ &    255\,cm$^{-1}$&  0.63\\
    dinitrobenzene & vibration& 1526\,cm$^{-1}$&     80\,cm$^{-1}$&    85\,cm$^{-1}$ &  0.06\\
    CdSe &  exciton & 1.82\,eV &   25\,meV &    29\,meV&0.02\\
    GaAs &  exciton & 1.52\,eV  &    3\,meV &  8\,meV &0.01\\\hline
    \end{tabular}
    \label{tab:verify}
\end{table*}

\subsection{Possible Consequences of USC in Solids}

The dipole lattice model predicts far-reaching consequences of ultrastrong LM coupling on the excited and ground state properties. We now show that its key characteristics may be identified qualitatively in entire classes of materials implying that USC likely shapes the ground and excited states of solids. 

\begin{figure*}
    \includegraphics[width=12cm]{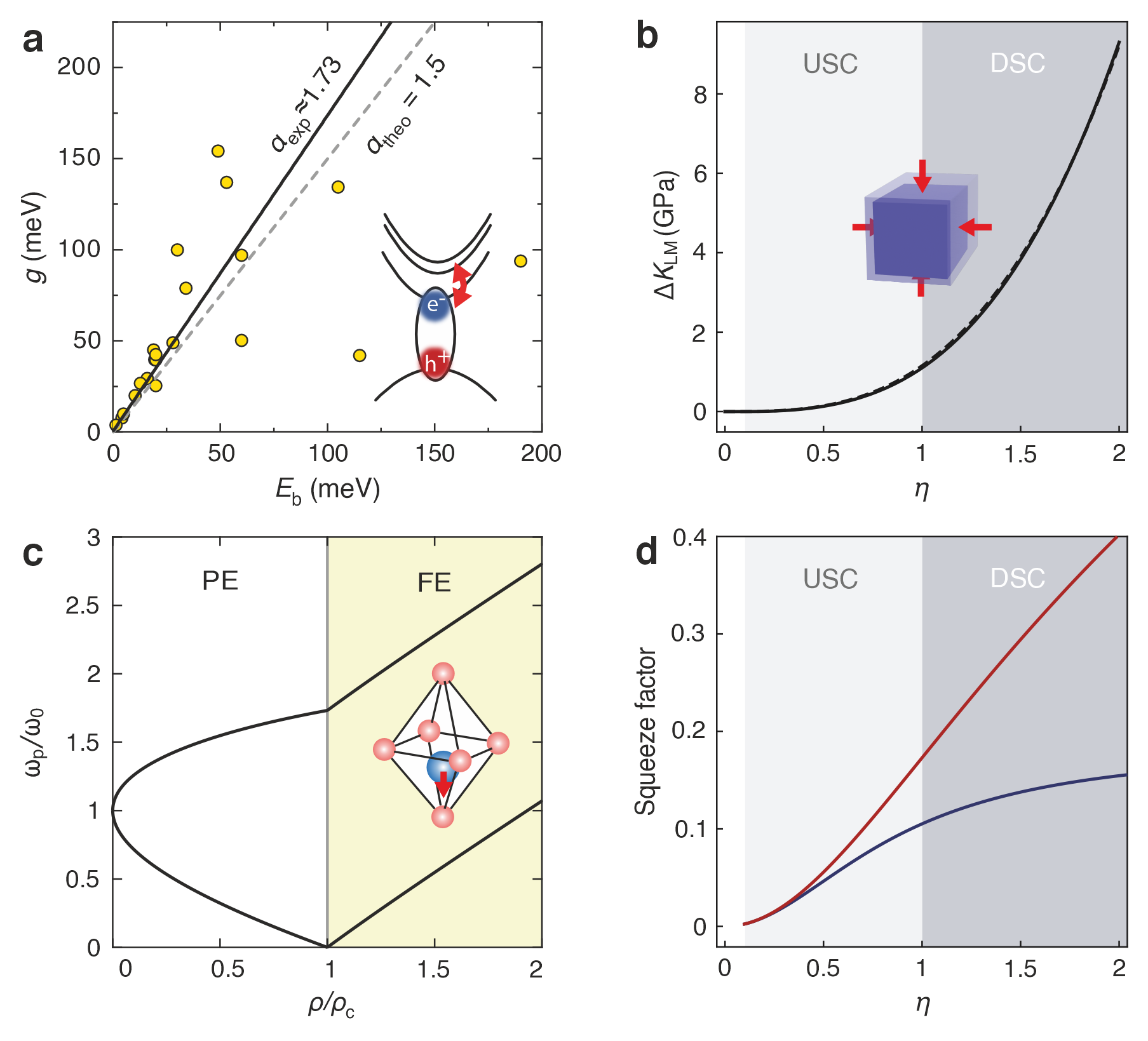}
    \caption{Material properties and ultrastrong LM coupling. (a) $g$ versus the exciton binding energy $E_b$ in semiconductors.  The lines show the predicted and measured mean $\alpha = g/E_b$, see labels. (b) LM  contribution to the bulk modulus as a function of normalized coupling strength, for $\hbar\omega_0 = 2\,$eV and $V_\mathrm{dip} = 0.1\,$nm$^3$. (c) Transition from paraelectric (PE) to ferroelectric (FE) phase. Polariton energies $\omega_\mathrm{P}$ are calculated at $\omega_\mathrm{pt,\mathbf k} = \omega_\mathrm{T}$. (d) Intrinsic squeezing of polariton states (at $\omega_\mathrm{pt,\mathbf k} = \omega_\mathrm{T}$) as function of normalized coupling strength for the lower (red) and upper (blue) polariton branches, see Methods.} 
    \label{fig:Fig4}
\end{figure*}

LM coupling influences excited-state properties beyond just the formation of polaritons. For instance, when the reduced coupling strength exceeds $\eta > 0.5$, light and matter decouple, leading to suppressed radiative decay.\cite{DeLiberato2014,Mueller2020} Additionally, when $g$ is comparable to the energetic separation between states, as is typical in the USC regime, excited states become strongly mixed. One example is phonon-polaritons in ferroelectrics, where each infrared-active transverse eigenmode couples to multiple longitudinal modes,\cite{Zhong1994} deviating from the conventional one-to-one correspondence between transverse and longitudinal phonons. Excitons, which exhibit a $1s, 2s, 3s, \dots$ series of optically active states separated by a fraction of the exciton binding energy $E_b$, are also susceptible to such mixing. This phenomenon has been proposed to alter the exciton binding energy and wavefunctions.\cite{Khurgin2001,Brodbeck2017} Using the hydrogen model for the exciton binding energy\cite{YuCardona} 
\begin{equation}
    E_b=\frac{me^4}{32\pi^2\epsilon_0^2\epsilon_\infty^2\hbar^2},
\end{equation}
$V_\mathrm{dip}=4\sqrt{2}r_{ex}^3$ with the Bohr radius $r_{ex}=4\pi\epsilon_0\epsilon_\infty\hbar^2/me^2$, and Eq.~\eqref{eq:g3D} we derive $g=1.5\sqrt f E_b$ for a dipole lattice of excitons. This implies that in 3D semiconductors ($f\approx 1$) the optically active exciton states are  separated by less than the LM coupling strength. Experimentally, we obtain an average $g = 1.7 E_b$ (with large variations) considering common semiconductors, Fig.~\ref{fig:Fig4}a and Supplementary Table~S7.
Excitonic states are likely mixed  in bulk semiconductors, highlighting the need for an exciton theory that incorporates LM coupling effects on binding energy and Bohr radius. Moreover,  mixing induced by LM coupling could also renormalize optically inactive excitons by coupling them to dipole-allowed transitions, as well as hybridize entire series of optically active states, including phonons, magnons, and their overtones.

The shift of the zero-point energy $\Delta E_\mathrm{ZP}$ in the USC regime changes the ground state of solids and their derived properties such as the structurally stable phase, phonons, and the mechanical response. For example, $\Delta E_\mathrm{ZP}$ adds a LM contribution to the bulk modulus, see Methods,
\begin{equation}
\label{eq:BulkLM}
\Delta K_\mathrm{LM}
= - V_\mathrm{dip}\frac{\partial^2 \Delta E_\mathrm{ZP}}{\partial V_\mathrm{dip}^2}
\approx 0.36 \frac{\hbar \omega_0}{V_\mathrm{dip}} \eta^3
\end{equation}
per material excitation (approximation for $V_\mathrm{dip}\ll \lambda_0^3$). This contribution  becomes appreciable if large $\eta$ and excitation frequencies are combined with small dipole volumes. For an excitation in the visible $\hbar\omega_0\approx 2\,$eV and a lattice constant 5\,{\AA} we find $\Delta K_\mathrm{LM}\approx 1\,$GPa, Fig.~\ref{fig:Fig4}b,  comparable to the bulk moduli of materials that  range 0.1-100\,GPa.

Increasing LM and dipole-dipole coupling progressively softens the lower polariton of the dipole lattice, ultimately driving a divergence in $\eta$. This effect  is evident in the large reduced coupling of phonon-polaritons in ferroelectrics, where $\omega_\mathrm{T}$ corresponds to a soft phonon, Fig.\ref{fig:mat}a blue squares, and in plasmonic supercrystals where the collective plasmon frequency $\omega_\mathrm{T}$ drops below $40\%$ of the single nanoparticle resonance $\omega_0$,\cite{Mueller2020} Fig.\ref{fig:mat}a pink circles. 
At a critical dipole density $\rho_c$ we predict an LM-induced phase transition that depends on the nature of the material excitation. In nanoparticle superlattices, a vanishing plasmon-polariton energy triggers an insulator-metal transition, changing the system from a Hopfield dielectric to a Drude metal. In phonon-polaritons, on the other hand, the finite nuclear displacement of the zero-frequency mode leads to the formation of permanent macroscopic dipoles. This transition occurs as the lower phonon polariton frequency approaches zero and becomes imaginary for $\rho > \rho_c$, Fig.\,\ref{fig:Fig4}c. A theory that incorporates non-linear terms via a Holstein-Primakoff mapping predicts that the lattice is driven into a ferroelectric phase with a macroscopic ground state polarization.\cite{Lamberto2024} The frequency of the lower polariton recovers as real-valued and increases with $\rho$, Fig.\,\ref{fig:Fig4}c and Methods, as observed for the soft mode frequency during the ferroelectric phase transition in perovskites.\cite{Luspin1980,Ashida2020,Ghosez1997} Finally, a vanishing frequency in excitonic systems enables the  spontaneous population of the ground state with electron-hole pairs without external excitations that may condense into the ground state of an excitonic insulator.\cite{Jerome1967,Ma2021}  This exotic state, analogous to Cooper pair condensation in superconductors, is expected to enable ultra-low-friction charge transport.

Beyond modifying material properties, USC reshapes the photonic states within bulk crystals, offering exciting prospects for quantum information and nanophotonics. The dipole lattice enables the coupling between photons of different energy, propagation direction, and polarization due to its translational symmetry   and the photon-photon interaction introduced by the $A^2$ term in the generalized Hopfield Hamiltonian.\cite{Barros2021} 
The presence of virtual excitations in the USC ground state leads to intrinsic squeezing of bulk polaritons, effectively suppressing quantum fluctuations of the bare excitations in the polaritonic ground state.\cite{Artoni1991,Quattropani2005,Hayashida2023} This squeezing is weak for $\eta \ll 1$ but becomes increasingly pronounced in the USC regime, Fig.\,\ref{fig:Fig4}d and Methods. As the system approaches DSC, the softening of the transverse mode enhances its virtual ground-state population, leading to a continuous increase in the intrinsic squeezing of the lower polariton,\cite{Hayashida2023} while the squeezing of the upper polariton saturates for $\eta > 1$. Although leveraging suppressed ground-state quantum fluctuation in polaritonic systems remains conceptually challenging, our findings suggest that the lower polariton in bulk materials with  ultrastrong or deep strong LM coupling provides an ideal platform for exploring these effects.

\section{Conclusion}

Ultrastrong LM coupling is a widespread phenomenon in solids, which naturally occurs for phonon-, exciton-, and plasmon-polaritons in various material platforms. Selected crystals even push into the DSC regime, where the coupling strength exceeds the excitation frequency. The fascinating USC physics -- excited state mixing, decoupling of light and matter, virtual excitations in the ground state, phase transitions, and squeezed photonic states -- should manifest in 3D systems and may even be hidden in plain sight under the known properties of crystals. The huge selection of bulk polaritons with USC collected in Fig.~\ref{fig:mat} has remained largely unexplored for polaritonic engineering and will serve as a guide for future material selection to study extreme LM coupling in solids. The generalized Hopfield Hamiltonian describes the USC physics of the dipole lattice in a cohesive way that, moreover, bridges to the existing framework of cavity QED. A  predictive implementation of this approach will have to include electron and nuclei, a direction that was   explored recently by the Pauli-Fierz quantum field theory.\cite{Ruggenthaler2023,Shin2021} 
The prevalence of USC in bulk materials opens exciting opportunities for practical applications. The soft-mode behavior of phonon-polaritons in ferroelectrics could be harnessed for tunable terahertz  devices, while the intrinsic squeezing of exciton polaritons may provide new pathways for room-temperature quantum light generation. The ability to control phase transitions via USC could be exploited in reconfigurable polaritonic materials. Future studies should explore whether USC can be actively tuned using nanostructuring, external fields, or strain engineering. Investigating nonlinear optical effects in the USC regime could reveal novel quantum phenomena beyond conventional polaritonics.



\section{Methods}

\subsection{Theory of the Transition Dipole Lattice}

We develop a microscopic model of delocalized excitations in solids and their interaction with photons. The excitations (phonons, excitons, plasmons) are represented by a lattice of localized transition dipoles. We consider a cubic lattice of these dipoles that couple with each other through Coulomb interaction as well as to a retarded photon field. Our theory is applicable to all dipole-active excitations in materials; it unifies existing approaches for specific material excitations, such as plasmon polaritons in metal supercrystals,\cite{Zhen2008,Lamowski2018,Fernique2020} and exciton- as well as phonon polaritons in molecular crystals.\cite{Johnson1976, DavydovBook, Barentzen1971} Below, we set up the theoretical framework to obtain analytic expressions for the polariton dispersion and collective lattice modes; further details can be found in the Supplementary Information.

We use several approximations to keep the theory accessible and easy to implement. We describe the excitations as dipolar and restrict our treatment to a single dipole per unit cell. We neglect Umklapp effects and cross-polarization coupling, which potentially change the polariton dispersion in ways that are negligible on the level of our fundamental description, see Refs.~\citenum{Zhen2008, Lamowski2018, Fernique2020, Johnson1976, DavydovBook}. None of the approximations pose a fundamental limit to our theory. It  may be implemented for other crystal lattices, higher-order modes, and may include interactions beyond the first Brillouin zone.\cite{Barros2021} 

The induced transition dipoles oscillate at frequency $\omega_0$ with dipole moment
\begin{equation}
    \mathbf{p}(\mathbf{R}) = - e f \sum_{\sigma = x, y, z} h_\sigma(\mathbf{R}) \hat{\boldsymbol \sigma},
\end{equation}
where $f$ is the oscillator strength of each dipole determined by the effective number of charges participating in the transition (see Suppl.\ Note S1.F.\ and G.\ and Suppl.\ Table S2). $h_\sigma(\mathbf{R})$ is the displacement field associated with the dipole at lattice position $\mathbf{R}$, and $\hat{\boldsymbol\sigma}$ a unit vector along a Cartesian coordinate. The dipoles are further characterized by the mass $m$ of the oscillating charges and are assumed to be three-fold degenerate. 

The dipoles couple through Coulomb interaction, with the pairwise interaction potential of two dipoles at lattice sites $\mathbf{R}$ and $\mathbf{R}'$
\begin{equation}
    V_\mathrm{d}^\mathrm{int}(\mathbf{R}, \mathbf{R}') = \frac{\mathbf{p}\cdot \mathbf{p}' - 3(\mathbf{p}\cdot \hat{\boldsymbol{\beta}})(\mathbf{p}'\cdot \hat{\boldsymbol{\beta}})}{4\pi \epsilon_0 \epsilon_\infty \beta^3},
\end{equation}
where $\beta = \vert \mathbf{R} - \mathbf{R}' \vert$ and $\hat{\boldsymbol{\beta}} = (\mathbf{R} - \mathbf{R}')/\beta$. $\epsilon_\infty$ is a background dielectric constant accounting for the screening by other optical transitions or a surrounding material.

Each dipole also couples to the transverse photon field with the vector potential
\begin{widetext}
\begin{equation}
    \label{eq:PhotonVectorPotential}
    A_\lambda(\mathbf{R}) = \sum_\mathbf{k} \sqrt{\frac{\hbar}{2 \epsilon_0 \epsilon_\infty N V_\mathrm{dip} \omega_{\mathrm{pt},\mathbf{k}}}} \left( a_{\mathbf{k}, \lambda} e^{i \mathbf{k}\cdot \mathbf{R}} + a_{\mathbf{k}, \lambda}^\dagger e^{-i \mathbf{k}\cdot \mathbf{R}} \right),
\end{equation}
\end{widetext}
where $\omega_{\mathrm{pt},\mathbf{k}} = c \vert \mathbf{k} \vert/\sqrt{\epsilon_\infty}$ is the photon dispersion. $a_{\mathbf{k}, \lambda}^\dagger$ and $a_{\mathbf{k}, \lambda}$ are creation and annihilation operators of a photon mode with wave vector $\mathbf{k}$ and polarization $\lambda$. $V_\mathrm{dip}$ is the volume of the primitive unit cell of the lattice and $N$ the number of unit cells. Light-matter coupling is described by the minimal coupling Hamiltonian as discussed in the Supplementary Text. 

The full Hamiltonian of the 3D lattice reads, see Supplementary Text,
\begin{equation}
\begin{split}
\mathcal{H} = &  \mathcal{H}_\mathrm{pt}^0 + \mathcal{H}_\mathrm{d}^0 + \mathcal{H}_\mathrm{d}^\mathrm{int} + \mathcal{H}_\mathrm{pt-d} 
\\ = & \sum_{\mathbf{k}, \lambda}  \hbar \omega_{\mathrm{pt},\mathbf{k}} a_{\mathbf{k}, \lambda}^\dagger  a_{\mathbf{k}, \lambda} + \sum_{\mathbf{k}, \sigma} \hbar \omega_0 b_{\mathbf{k}, \sigma}^\dagger  b_{\mathbf{k}, \sigma}
\\ & + \sum_{\mathbf{k}, \sigma, \sigma'} \hbar d s_{ \mathbf{k}, \sigma, \sigma'}(b_{\mathbf{k}, \sigma}^\dagger + b_{-\mathbf{k}, \sigma})(b_{-\mathbf{k}, \sigma'}^\dagger + b_{\mathbf{k}, \sigma'})
\\ & + \sum_{\mathbf{k}, \lambda, \sigma} i \hbar g \xi_\mathbf{k} (a_{\mathbf{k}, \lambda}^\dagger + a_{-\mathbf{k}, \lambda})(b_{-\mathbf{k}, \sigma}^\dagger - b_{\mathbf{k}, \sigma})
\\ & + \sum_{\mathbf{k}, \lambda} \hbar d \xi_\mathbf{k}^2 (a_{\mathbf{k}, \lambda}^\dagger + a_{-\mathbf{k}, \lambda})(a_{-\mathbf{k}, \lambda}^\dagger + a_{\mathbf{k}, \lambda}),
\end{split}
\end{equation}
with $b_{\mathbf{k}, \sigma}^\dagger$ and $b_{\mathbf{k}, \sigma}$ bosonic creation and annihilation operators of dipole excitations with wave vector $\mathbf{k}$ and polarization $\sigma$. 

\begin{equation}
    d = \frac{e^2 f}{4\epsilon_0 \epsilon_\infty \omega_0 m V_\mathrm{dip}}
\end{equation}
is the dipole coupling constant and\cite{Lamowski2018}
\begin{equation}
\label{eq:DipoleSum3D}
s_{ \mathbf{k}, \sigma, \sigma'}= \sum_{\boldsymbol\beta} \cos{(\boldsymbol\beta \cdot \mathbf{k})} \frac{\delta_{\sigma, \sigma'} - 3(\hat{\boldsymbol \sigma} \cdot \hat{\boldsymbol\beta})(\hat{\boldsymbol \sigma}' \cdot \hat{\boldsymbol\beta})}{2\pi \rho^3/V_\mathrm{dip}}
\end{equation}
a lattice sum that contains all information about the structure of the dipole lattice.

\begin{equation}
\label{eq:g3D_methods}
    g = \sqrt{\frac{e^2 f}{4\epsilon_0 \epsilon_\infty m V_\mathrm{dip}}} = \sqrt{d \omega_0}
\end{equation}
is the LM  coupling strength and $\xi_\mathbf{k} = \sqrt{\omega_0 / \omega_\mathrm{pt}(\mathbf{k})}$. 

To estimate upper bounds for $g$ in materials we considered the excitation of an oscillating electron ($f=1, m=m_e$) in the absence of dielectric screening. Equation~\eqref{eq:g3D_methods} simplifies to $g=0.6\,$eV\,nm$^{3/2}/\sqrt{V_\mathrm{dip}}$. A characteristic lattice constant in the periodic table is 5\,{\AA} yielding a primitive fcc cell volume of $0.03\,$nm$^3$ or a sphere volume $0.02$\,nm$^3$ and thus $g\approx 3\,$eV. The LM coupling strength for a single dipole in a photonic cavity $g_\mathrm{cav}$ is obtained from Eq.~\eqref{eq:g3D_methods} when replacing $V_\mathrm{dip}$ by $V_\mathrm{cav}$, where $V_\mathrm{cav}$ is the cavity mode volume. To obtain an upper bound we first consider a planar photonic cavity with a minimum mode volume  $(\lambda/2)^3$. For visible excitations ($\lambda\sim 500\,$nm) the LM coupling strength is then limited to 0.1\,meV. Sub-wavelength plasmonic cavities reach 10\,meV with a theoretically predicted maximum of $1\,$eV, but at the expense of plasmonic cavity losses $>100\,$meV.\cite{Roelli2016,Barbry2015} USC in cavity-based experiments is reached by placing several $N$ dipoles into the cavity, increasing the LM coupling strength to $\sqrt N g_\mathrm{cav}$. However, as long as the volume occupied by the dipoles is smaller than the cavity volume, LM coupling by particles in a cavity remains below the limit set by the corresponding bulk material.

Neglecting the terms of the Hamiltonian that contain photon operators, we obtain the collective dipole dispersion, derivation in Supplementary Text,
\begin{equation}
\label{eq:collective_frequency_3D}
\omega_{\mathrm{coll}, \mathbf{k}, \sigma}
= 
\sqrt{\omega_0^2 + 2\omega_0 d s_{ \mathbf{k}, \sigma}}.
\end{equation}
It describes  transverse (T) and longitudinal (L) collective dipole excitations shown by blue lines in Fig.~\ref{fig:Fig3}b, i.e.\ $\omega_{\mathrm{coll}, \mathbf{k}, \sigma} = \omega_{\mathrm{T}, \mathbf{k}, \sigma}$ for $\hat{\boldsymbol\sigma} \perp \mathbf{k}$ and $\omega_{\mathrm{coll}, \mathbf{k}, \sigma} = \omega_{\mathrm{L}, \mathbf{k}, \sigma}$ for $\hat{\boldsymbol\sigma} \parallel \mathbf{k}$. Close to the $\Gamma$ point, the dipole sums become unstable and do not converge. For small wavevectors the frequencies in an fcc lattice may be approximated  by $\omega_\mathrm{T} \approx \sqrt{\omega_0^2 - 4d \omega_0/3}$ and $\omega_\mathrm{L} \approx \sqrt{\omega_0^2 + 8d \omega_0/3}$.\cite{Cohen1955} Corresponding expressions exist for the other cubic lattices, including $s_{0,\mathrm T} = -\sqrt{2}/3$, $s_{ 0,\mathrm L} = 2\sqrt{2}/3$ for simple cubic and $s_{0,\mathrm T} = -\sqrt{3}/(2\sqrt{2})$, $s_{  0, \mathrm L} = \sqrt{3/2}$ for body centered cubic.\cite{Cohen1955,Lamowski2018,Barros2021}

Diagonalizing the full Hamiltonian, see Supplementary Text, leads to the polariton dispersion
\begin{widetext}
\begin{equation}
\label{eq:3DpolaritonDispersion}
\omega_{\mathbf{k},\sigma, \pm} = 
\sqrt{\frac{\omega_{\mathrm{pt}, \mathbf{k}}^2 + \omega_{\mathrm{T}, \mathbf{k}, \sigma}^2 + 4 g^2
\pm
\sqrt{(\omega_{\mathrm{pt}, \mathbf{k}}^2 +  \omega_{\mathrm{T}, \mathbf{k}, \sigma}^2 +  4 g^2)^2 - 4 \omega_{\mathrm{pt}, \mathbf{k}}^2 \omega_{\mathrm{T}, \mathbf{k}, \sigma}^2}
}{2}},
\end{equation}
\end{widetext}
with the index $+$ for the upper and $-$ for the lower polariton branch. This dispersion is compared to a dispersion obtained from numerical simulations solving Maxwell's equations in the Supplementary Information finding excellent agreement for the example of a dipole lattice of Lorentz oscillators.

\subsection{Ground State Energy and Bulk Modulus}

The change in ground state energy, compared to a material without LM  coupling, is calculated from the energetic difference between the polariton energies and the energies of the uncoupled components, Eq.\,(\ref{eq:BGE0}).\cite{Ciuti2005}
We sum over all possible polarizations $\sigma$ and average over $N_k$ wavevectors $\mathbf{k}$ in the first Brillouin zone. The prefactors for the uncoupled excitations account for the three-fold degeneracy of the dipoles and two-fold degeneracy of the photons. For ease of implementation, we restrict the summation to the irreducible part of the Brillouin zone, Suppl.\ Fig.\,S4. 

To evaluate $\Delta E_\mathrm{ZP}$ numerically, we introduce a photon energy cutoff. It is motivated by the distribution of charges in real space;  photons with wavelengths  shorter than the width of the charge distribution cannot couple to charges and are discarded in $\Delta E_\mathrm{ZP}$. We set the minimum photon wavelength that couples to the dipoles to $\lambda_\mathrm{min}=a=V_\mathrm{dip}^{1/3}$. In discussing $\Delta E_\mathrm{ZP}$ this lattice constant $a$ is compared to the vacuum wavelength of the dipole frequency $\lambda_0 = 2\pi c_0/\omega_0$. 

For $a\ll \lambda_0$, the anti-crossing of the two polariton branches occurs close to the Brillouin zone center. Most natural crystals belong to this regime. In this case, Coulomb dipole-dipole interaction dominates the energy scale across most of the Brillouin zone. The change in  zero point energy is approximated as 
\begin{equation}
\label{eq:ZPEnergyCoulomb}
\Delta E_\mathrm{ZP} \approx 
(2\hbar \omega_{\mathrm{T},\mathbf{k}} + \hbar \omega_{\mathrm{L},\mathbf{k}}-  3 \hbar \omega_0)/2 = 
\frac{3}{2}\omega_0\left(\frac{2+\sqrt{1+4\eta^2}}{\sqrt{9+12\eta^2}} - 1\right),
\end{equation}
where we used $\omega_\mathrm{T} \approx \sqrt{\omega_0^2 - 4g^2/3}$ and $g = \omega_0/\sqrt{1/\eta^2+4/3}$.

The zero point energy in Eq.~\eqref{eq:BGE0} depends on the dipole density and thus the dipole volume $V_\mathrm{dip}$. This implies  that LM  coupling contributes to the bulk modulus $K$ of materials. Using Eqs.\,(\ref{eq:g3D}) and (\ref{eq:ZPEnergyCoulomb}), we obtain a change in the bulk modulus due to  dipole-dipole coupling
\begin{equation}
\label{eq:BulkLM_methods}
\Delta K_\mathrm{LM}
= - V_\mathrm{dip}\frac{\partial^2 \Delta E_\mathrm{GS}}{\partial V_\mathrm{dip}^2}
= 
\frac{4 \hbar \omega_0 \eta^4}{3V_\mathrm{dip}} \cdot
\frac{1+2/\sqrt{(1+4\eta^2)^3}}{\sqrt{9+12\eta^2}} 
\approx
0.36\, \frac{ \hbar \omega_0 \eta^3}{V_\mathrm{dip}}.  
\end{equation}
The approximation was obtained from a fit to the full expression. The solid line in Fig.~\ref{fig:Fig4}b shows the full expression and the dashed line the approximation. 

\subsection{Phase Transition}

The LM induced phase transition occurs at the coupling strength where $\omega_\mathrm{T}\rightarrow 0$, which is given by $g_c=\sqrt3 \omega_0/2$ for fcc. This implies a critical dipole density 
\begin{equation}
    \rho_c=\frac{3\omega_0^3}{e^2\epsilon_0\epsilon_\infty m}
\end{equation}
for the occurance of a phase transition in an fcc dipole lattice. 

To study polaritons across the LM phase transition we used the gauge-equivalent lattice model extended to include condensation as proposed by Lamberto~\emph{et al.}, Ref.\,\citenum{Lamberto2024}. It yields a generalized polariton dispersion 
\begin{equation}
\label{eq:pol_phase}
\omega_{\mathbf{k},\sigma, \pm} = 
\sqrt{\frac{\omega_{\mathrm{pt}, \mathbf{k}}^2 + \omega_{\mathrm{L}, \mathbf{k}, \sigma}^2
\pm
\sqrt{(\omega_{\mathrm{pt}, \mathbf{k}}^2 +  \omega_{\mathrm{L}, \mathbf{k}, \sigma}^2)^2 - 4 \omega_{\mathrm{pt}, \mathbf{k}}^2 \omega_{\mathrm{T}, \mathbf{k}, \sigma}^2}
}{2}}.
\end{equation}
where the longitudinal and transverse frequencies in each phase can be written as
\begin{equation}
\label{eq:TransverseFreqParaelectricFerroelectric}
\omega_\mathrm{T,\mathbf{k}}=\left\{\begin{split}
\sqrt{\omega_0^2+2g^2s_{\mathbf{k},\mathrm T}} &\mathrm{,~~for~~} g<\sqrt{\omega_0^2/|s_{\mathbf k,\mathrm T}|}  \\\sqrt{4g^4s_{\mathbf k,\mathrm T}^2/\omega_0^2-\omega_0^2}
&\mathrm{,~~for~~} g>\sqrt{\omega_0^2/|s_{\mathbf k,\mathrm T}|} 
\end{split}\right. ,
\end{equation} and 
\begin{equation}
\label{eq:LongitudinalFreqParaelectricFerroelectric}
\omega_\mathrm{L,\mathbf k}=\left\{\begin{split}
\sqrt{\omega_0^2+2g^2(2+s_{\mathbf k,\mathrm T})} &\mathrm{,~~for~~} g<\sqrt{\omega_0^2/|s_{\mathbf k,\mathrm T}|}  \\\sqrt{4g^4s_{\mathbf k,\mathrm T}^2/\omega_0^2-\omega_0^2-2\omega_0^2/s_{\mathbf k,\mathrm T}}
&\mathrm{,~~for~~} g>\sqrt{\omega_0^2/|s_{\mathbf k,\mathrm T}|} 
\end{split}\right. .
\end{equation}
The polariton frequencies in Fig.~\ref{fig:Fig4}c and Eqs.~\eqref{eq:TransverseFreqParaelectricFerroelectric} and \eqref{eq:LongitudinalFreqParaelectricFerroelectric} were calculated  by evaluating Eq.~\eqref{eq:pol_phase} under the condition $\omega_{\mathrm{pt},\mathbf k}=\omega_{\mathrm T,0}$. Equation~\eqref{eq:pol_phase} is identical to Eq.~\eqref{eq:3DpolaritonDispersion} for  $g<\sqrt{\omega_0^2/|s_{\mathbf k,\mathrm T}|} $.

\subsection{Intrinsic Squeezing}

In the polariton Hamiltonian, the  forward ($\mathbf k$) and backward ($-\mathbf k)$ moving bare excitations are mixed through the counter-rotating terms ($a_\mathbf{k}a_{-\mathbf k}$, $a_\mathbf{k}b_{-\mathbf k}$, $b_\mathbf{k}b_{-\mathbf k}$ and h.c.). This causes the polaritons to be intrinsically squeezed states of the bare photon and dipole excitations.  The intrinsic squeezing of the polariton states is characterized by the squeeze factor \cite{Artoni1991}
\begin{equation}
\label{eq:PolaritonSqueezing}
 r_{\mathbf k,\sigma,\pm}=\tanh^{-1}\left(\frac{|y_{\mathbf k,\sigma,\pm}|^2+|z_{\mathbf k,\sigma,\pm}|^2}{1+|y_{\mathbf k,\sigma,\pm}|^2+|z_{\mathbf k,\sigma,\pm}|^2}\right),
\end{equation}
where $y_{\mathbf k,\sigma,\pm}$ and $z_{\mathbf k,\sigma,\pm}$ are the contributions of the backward propagating bare excitations to the polariton operator
\begin{equation}
    \label{eq:PolaritonOperator}
    p_{\mathbf{k},\sigma, \pm} = w_{\mathbf{k}, \sigma, \pm} b_{\mathbf{k},\sigma} + x_{\mathbf{k}, \sigma, \pm} a_{\mathbf{k},\sigma} + y_{\mathbf{k}, \sigma, \pm} b_{-\mathbf{k},\sigma}^\dagger + z_{\mathbf{k}, \sigma, \pm} a_{-\mathbf{k},\sigma}^\dagger.
\end{equation}
This factor governs both the non-Poissonian photon distribution in the ground state and the oscillation of the quantum noise for different bare excitation quadratures.\cite{Artoni1991} Fig.~\ref{fig:Fig4}d plots $r_{\mathbf k,\sigma,-}$ (red) and $r_{\mathbf k,\sigma,+}$ (blue) at $\mathbf k$ where $\omega_{\mathrm{pt},\mathbf k}=\omega_{\mathrm T}$.

\subsection{Evaluating LM Coupling from Experimental Data}

The LM  coupling strength $g$ of 3D materials was determined from the collective transverse and longitudinal frequencies observed via optical spectroscopy. Using Eq.~\eqref{eq:collective_frequency_3D} we obtain 
\begin{equation}
g = \frac{\sqrt{\omega_\mathrm{L}^2 - \omega_\mathrm{T}^2}}{2}.
\end{equation}
For preparing Fig.~\ref{fig:mat} we identified $\omega_\mathrm T = \omega_m$. Some experiments reported the high-frequency $\epsilon(\infty)$ and low-frequency $\epsilon(0)$ dielectric constants together with $\omega_\mathrm{T}$ or $\omega_\mathrm{L}$. These quantities are connected by the Lyddane-Sachs-Teller relation \cite{YuCardona}
\begin{equation}
    \frac{\omega_\mathrm{L}^2}{\omega_\mathrm{T}^2} = \frac{\epsilon(0)}{\epsilon(\infty)}
\end{equation}
or
\begin{equation}
g = \frac{\omega_\mathrm{T}}{2}\sqrt{\frac{\epsilon(0)}{\epsilon(\infty)} - 1}.
\end{equation}

Cavity experiments of strong and ultrastrong LM coupling directly report $g$ and $\omega_m$. Some studies used the Rabi \emph{splitting} $2g$ in the definition of the reduced coupling strength. In these cases we  report the reduced coupling according to our definition, i.e., using the Rabi \emph{frequency} or LM coupling strength ($\eta = g/\omega_\mathrm T).$

We provide Supplementary Tables~S3-S16
that collect the absolute LM  coupling strength $g$, the transverse dipole frequency $\omega_\mathrm{T}=\omega_m$, the reduced coupling $\eta = g/\omega_\mathrm{T}$, and the dielectric screening $\epsilon_\infty$ (where available) for all materials and cavity setups. The data was used to prepare Fig.~\ref{fig:mat}.

\textbf{Data Availability:} All data is available in the manuscript or the Supplementary Information. In addition, we deposited the data of Fig.~\ref{fig:mat}a and b in the repository ReFUbium under doi XXXX (doi will be included at proof stage), see also Supplementary Tables S3-S16.

\bibliographystyle{naturemag}
\bibliography{LB.bib}

\section{Acknowledgements}
The authors thank C.\ Thomsen for a critical reading of the manuscript and useful discussions. This work was supported by the European Research Council (ERC) under grant DarkSERS-772\,108, the German Research Foundation (DFG) under grant 504\,656\,879, the Berlin Center for Global Engagement (BCGE), the Center for International Collaboration (CIC) as well as the SupraFAB Research Center at Freie Universit\"at Berlin. N.S.M.\ thanks A.\ Paarmann and M.\ Spencer for useful discussions and acknowledges funding from the Deutsche Forschungsgemeinschaft (DFG, German Research Foundation) - Projektnummer 551280726, as well as support from the German National Academy of Sciences Leopoldina through the Leopoldina Postdoc Scholarship. E.B.B.\ thanks Technische Universit\"at Berlin for hospitality during the preparation of this manuscript and acknowledges support from 
CNPq, and CAPES.

\section{Author contributions}
All three authors contributed in multiple ways to the design, research, and presentation of this study.

\section{Author information}
The authors declare no competing interests. Correspondence and requests for materials should be addressed to reich@physik.fu-berlin.de.

\end{document}